\documentclass[a4paper,twocolumn,nofootinbib,showpacs,aps,floatfix,superscriptaddress]{revtex4}
\usepackage{graphicx}
\usepackage{amsmath}
\usepackage{bm}

\newcommand {\fabs}[1] {\left| #1 \right|}
\newcommand{\mathbb}[1]{}
\begin{document}
\title{Quantum catcher - stopping particles of unknown velocities} 

\author{S. Schmidt}
\email{soenke.schmidt@tu-bs.de}
\affiliation{Institut f\"ur Mathematische Physik, TU Braunschweig, Mendelssohnstrasse 3, 38106 Braunschweig, Germany}
\author{J. G. Muga}
\email{jg.muga@ehu.es}
\affiliation{Departamento de Qu\'\i mica F\'\i sica, UPV-EHU, Apartado 644, 48080 Bilbao, Spain}
\author{A. Ruschhaupt}
\email{a.ruschhaupt@tu-bs.de}
\affiliation{Institut f\"ur Mathematische Physik, TU Braunschweig, Mendelssohnstrasse 3, 38106 Braunschweig, Germany}
\begin{abstract}
We propose a method to stop particles of unknown velocities by
collision with an accelerated wall with trajectory $\sim \sqrt{t}$. We
present classical and quantum mechanical descriptions
and numerical simulations that show the efficiency of the method.
\end{abstract}
\pacs{
42.50.-p 
}
\maketitle
%
%
%
\section{Introduction}
A beam of particles can be slowed down by reflecting them from
a potential wall (or ``mirror'') moving in the
same direction. An early example is the production of an
ultracold beam of neutrons colliding with a moving
Ni-surface \cite{neutrons}. Moving mirrors for cold atom waves have
been also implemented with a time-modulated, blue-detuned evanescent
light wave propagating along the surface of a glass prism
\cite{Dalibard1,Dalibard2}. More recently, beams of Helium
atoms have been slowed down using a Si-crystal
on a spinning rotor \cite{Raizen06,m07}. Also, Rb atoms have been stopped 
with a moving
magnetic mirror on a conveyor belt \cite{Guery06}, which provides a
promising mechanism to generate a continuous, intense and slow
beam of atoms.
There is clearly a great potential for practical applications of such 
processes and much interest in their fundamental properties and optimization.
In most cases the analysis is made with classical trajectories,
but quantum motion effects may become important for ultracold atoms, as shown   
in a recent study on matter-wave/moving-mirror interaction \cite{Campo}. One more limitation of standard settings so far is  
that the mirror motion is adapted only to stop a specific initial beam velocity.   
We may however consider the much more general case in which the initial velocities    
are broadly distributed, or that they are unknown. Stopping a beam in these 
conditions is a much more challenging objective, and the central subject of the present paper.  

Can we stop a classical particle of unknown velocity  
with a moving hard wall? The answer is yes and the trajectory 
of the wall is surprisingly simple, as we shall see in Sect. \ref{sect2}. 
We shall also explore in Sect. \ref{sect3} the extent to which the same
methodology can be applied in the quantum case, 
and provide numerical examples.  
\section {Classical ensemble of non-interacting particles\label{sect2}}
\begin{figure}[t]
\begin{center}
\includegraphics[angle=0,width=0.90\linewidth]{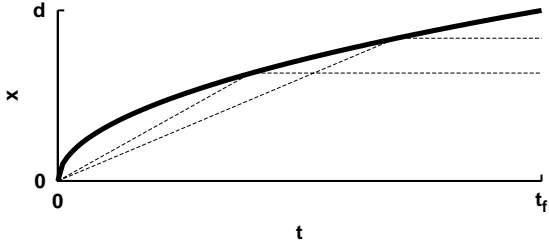}
\end{center}
\caption{\label{fig1} Scheme of the stopping of classical particles: a hard wall moving with trajectory
$x_m (t) = d \sqrt{t/t_f}$ (solid line); examples of two particle trajectories
with different initial velocities (dashed lines).}
\end{figure}

Let us start by assuming a 
classical point-particle emitted at $x=0$, $t=0$, moving with unknown velocity $v\geq 0$ along $x$. 
A heavy hard wall (compared to the mass of the particle) moves also with trajectory $x_m (t)$ and initial condition $x_m(0)=0$. 
If the particle touches the wall at time $t$ with velocity $v_s$ and the wall is
moving with velocity $v_m$ at $t$, then the 
final velocity $v_f$ of the particle after the perfect reflection is
\begin{eqnarray}
v_f = -v_s + 2 v_m.
\end{eqnarray}
This can be easily seen in the reference frame in which the
mirror is at rest at time $t$.
To fulfill the goal that the particle is at rest in the laboratory frame 
after the collision, the mirror velocity should be 
$v_m = v_s/2$ at the time of the collision.
Since the trajectory of the particle with velocity $v>0$ is $x(t) = v t$ and the
trajectory of the mirror is $x_m (t)$, the time $t_c$ of the collision is
given as a solution of $x_m (t_c) = v t_c$. The condition for stopping the
particle is now
\begin{eqnarray*}
\frac{d x_m}{dt} (t_c) = \frac{v}{2}.
\end{eqnarray*}
Using $v = x_m (t_c)/t_c$ we get
\begin{eqnarray*}
\frac{d x_m}{dt} (t_c) = \frac{x_m (t_c)}{2t_c}.
\end{eqnarray*}
The trajectory for the mirror we are looking for is found as the solution of
this ordinary differential equation with the initial condition $x_m(0) = 0$, 
\begin{eqnarray}
x_m(t) = \alpha \sqrt{t},
\label{solution}
\end{eqnarray}
where $\alpha > 0$ is in principle arbitrary, although its value will determine the
location and time of the particle-mirror collision, which is quite important
in a practical implementation with limited space and time.   
Such a moving wall will stop particles starting at the origin and moving with
positive velocity independent of their initial velocity, see also
Fig. \ref{fig1}. Instead of $\alpha$ we will deal with more intuitive
parameters: we assume a final time $t_f$, this may be a maximal time we are
ready to consider in our experiment. At
this time, the wall has moved a distance $d$ (see Fig. \ref{fig1}), so 
we get $\alpha = d/\sqrt{t_f}$. Another important quantify is $v_b := d/t_f$,
the boundary velocity for no collision until $t_f$, i.e., a particle with an
initial velocity $v_s < v_b$ will not hit the mirror until $t_f$.

\begin{figure}[t]
\begin{center}
\includegraphics[angle=0,width=0.90\linewidth]{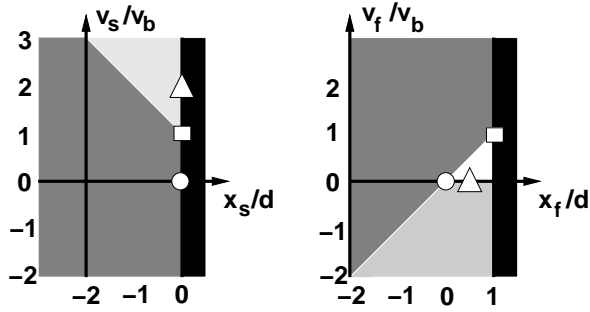}
\end{center}
\caption{\label{fig2} Scheme of the transformation between initial parameters
and final parameters, the wall is indicated by the black box,
the dark gray region is for free motion, and the light gray region for motion with
collision. 
The symbols provide examples connecting final and initial parameters: $x_s/d = 0, v_s/v_b = 0 \to x_f/d = 0, v_f/v_b=0$ (free motion, circles);
$x_s/d = 0, v_s/v_b = 1 \to x_f/d=1, v_f/v_b = 1$ (free motion, boxes);
$x_s/d = 0, v_s/v_b = 2 \to x_f/d=1/2, v_f/v_b = 0$ (collision, triangles).}
\end{figure}

\begin{figure}[t]
\begin{center}
\includegraphics[angle=0,width=0.90\linewidth]{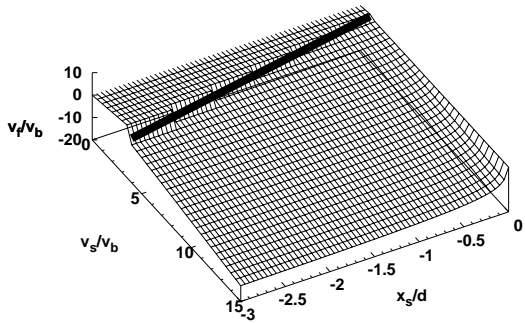}
\end{center}
\caption{\label{fig3} Final velocity $v_f (x_s, v_s)$, the line
  at $v_s/v_b=1-x_s/d$ separates the regions with or without a collision
  before $t_f$.} 
\end{figure}

Let us now examine the effect of the mirror with trajectory $x_m(t) = d
\sqrt{t/t_f}$ in more detail, allowing for a more general
scenario where the initial position of the 
particle is not necessarily zero but $x(0)=x_s \le 0$. We are interested
in the particle's position $x_f$ and velocity $v_f$ at $t_f$.
It is useful to introduce dimensionless variables to simplify the notation
in the following, namely
\begin{eqnarray}
\chi = \frac{x}{d} \, , \,
\nu  = \frac{v}{v_b} \, , \, \tau = \frac{t}{t_{f}}.
\end{eqnarray}
By a straightforward calculation we get for the position and velocity of the
particle at time $\tau = 1$ for $\chi_s \le 0$:
\begin{widetext}
\begin{eqnarray*}
\nu_f (\chi_s, \nu_s) &=& \left\{\begin{array}{ll}
\frac{1}{\eta(\chi_s, \nu_s)} - \nu_s & :\, \nu_s > 1 - \chi_s\\
\nu_s & : \, \mbox{otherwise}
\end{array}
\right.\\
\chi_f (\chi_s, \nu_s) &=& \left\{\begin{array}{ll}
\chi_s + \frac{1}{\eta(\chi_s, \nu_s)} - \nu_s + 2 \nu_s \eta^2(\chi_s, \nu_s)
- \eta(\chi_s, \nu_s)
& :\, \nu_s > 1 - \chi_s\\
\chi_s + \nu_s & : \, \mbox{otherwise}
\end{array}
\right.,
\end{eqnarray*}
with
\begin{eqnarray}
\eta(\chi_s, \nu_s) = \frac{1}{2 \nu_s}
        \left(1 + \sqrt{1-4 \chi_s \nu_s}\right).
\end{eqnarray}
If $\nu_s < 1 - \chi_s$ 
the particle keeps its initial velocity, i.e., it moves too slowly to 
collide with the moving-wall before $\tau_1$, see also Fig. \ref{fig2}. 
We can also work out the inverse transformation to get the initial position and
velocity from the final parameters for $\chi_f < 1$:
\begin{eqnarray*}
\nu_s (\chi_f, \nu_f) &=& \left\{\begin{array}{ll}
\lambda (\chi_f, \nu_f) - \nu_f & :\, \chi_f > \nu_f \;\mbox{and}\; \nu_f\le 0\\
\nu_f & : \, \mbox{otherwise}
\end{array}
\right.\\
\chi_s (\chi_f, \nu_f) &=& \left\{\begin{array}{ll}
\chi_f - \nu_f + \frac{2}{\lambda^2(\chi_f, \nu_f)} \left(\nu_f - \frac{\lambda(\chi_f,
  \nu_f)}{2} \right)& :\, \chi_f > \nu_f \;\mbox{and}\; \nu_f \le 0\\
\chi_f - \nu_f & : \, \mbox{otherwise}
\end{array}
\right.,
\end{eqnarray*}
with 
\begin{eqnarray}
\lambda(\chi_f, \nu_f) = \frac{1}{2 (\chi_f - \nu_f)} \left(1 + \sqrt{1-4 \nu_f (- \nu_f + \chi_f)} \right).
\end{eqnarray}
\end{widetext}
The function $v_f (x_s, v_s) = v_b \nu_f (x_s/d, v_s/v_b)
= v_b \nu_f (\chi_s, \nu_s)$ is important as it provides the extent to which 
the moving mirror fails to stop the particles 
when they deviate from the ideal conditions $\chi_s =
0$ and $\nu_s \ge 1$. This function is shown in Fig. \ref{fig3}.
Let us examine the case $\nu_s > 1 - \chi_s$ (and $\chi_s \le 0$), i.e. a collision has occured before $\tau=1$: 
the final velocity is negative, $\nu_f \le 0$; also
\begin{eqnarray*}
\frac{\partial \nu_f}{\partial \nu_s} &=& -1 - \frac{1}{2 \chi_s}
\left(\frac{1}{4 \chi_s^2} - \frac{\nu_s}{\chi_s}\right)^{-\frac{1}{2}}\\
&=& - 1 + \frac{1}{\sqrt{1 - 4 \nu_s \chi_s}}\\
&<& 0, 
\end{eqnarray*} 
and
\begin{eqnarray*}
\frac{\partial \nu_f}{\partial \chi_s} 
&=& \frac{-\frac{1}{2 \chi_s^3} + \frac{\nu_s}{\chi_s^2}}{2 \sqrt{\frac{1}{4 \chi_s^2} -
    \frac{\nu_s}{\chi_s}}} - \frac{1}{2 \chi_s^2}\\
&=& \frac{1}{2 \chi_s^2} \left(\frac{-\frac{1}{2 \chi_s} + \nu_s}{\sqrt{\frac{1}{4 \chi_s^2} -
    \frac{\nu_s}{\chi_s}}} - 1 \right)\\
&\ge& 0,
\end{eqnarray*}
so the absolute value of the final velocity is increasing with increasing
$\fabs{\chi_s}$ and $\fabs{\nu_s}$.

Fig. \ref{fig3} suggests to us a possible strategy to select the parameters
$d$ and $t_f$ and optimize the stopping: assume
that the initial parameters $x_s$ and
$v_s$ are fixed 
(they may correspond to estimates of the least favorable 
values expected or permitted, such as the farthest distance from the origin allowed by the initial geometry of the launching conditions and a lower bound for the speed).
First we choose
$d$ and $t_f$ such that
\begin{eqnarray*}
v_b = \frac{d}{t_f} \ll v_s.
\end{eqnarray*}
We have for the final velocity
\begin{eqnarray*}
v_f (x_s, v_s) = v_b \nu_f \left(\frac{x_s}{d}, \frac{v_s}{v_b}\right)
\stackrel{d\to\infty}{\longrightarrow} 0\;\quad (v_b\, \mbox{fixed}).
\end{eqnarray*}
According to Fig. \ref{fig3}, this means that if we 
make $d$ large enough, increasing also $t_f$ so that $v_b$ remains constant, the final velocity goes to zero, even with imperfect 
conditions. 
The condition $v_b (1- x_s/d) < v_s$ should be also satisfied for the chosen $d$,
so that the particle
collides with the wall, but this is not a problem because
\begin{eqnarray*}
v_b \left(1- \frac{x_s}{d}\right)
\stackrel{d\to\infty}{\longrightarrow}
v_b \ll v_s\;\quad (v_b\, \mbox{fixed}).
\end{eqnarray*}

In the following we shall discuss the more general case in which 
the initial position and the velocity of the particle
are characterized by a probability density $p_s (x_s, v_s)$.  
The final probability density for position and velocity of the
particle is given by
\begin{eqnarray*}
p_s (x_f, v_f) = p_s [x_s (x_f, x_f), v_s (v_f, v_f)].
\end{eqnarray*}
In particular we shall consider 
examples with
\begin{eqnarray*}
p_s (x, v) = \frac{1}{N} \exp\left[-\frac{(v-v_0)^2}{2\Delta v^2}
- \frac{(x-x_0)^2}{2\Delta x^2}\right]
\end{eqnarray*}
for $x < 0$, and $p_s (x,v) = 0$ for $x \ge 0$ ($N$ being
a normalization constant).

\begin{figure}[t]
\begin{center}
\includegraphics[angle=0,width=0.90\linewidth]{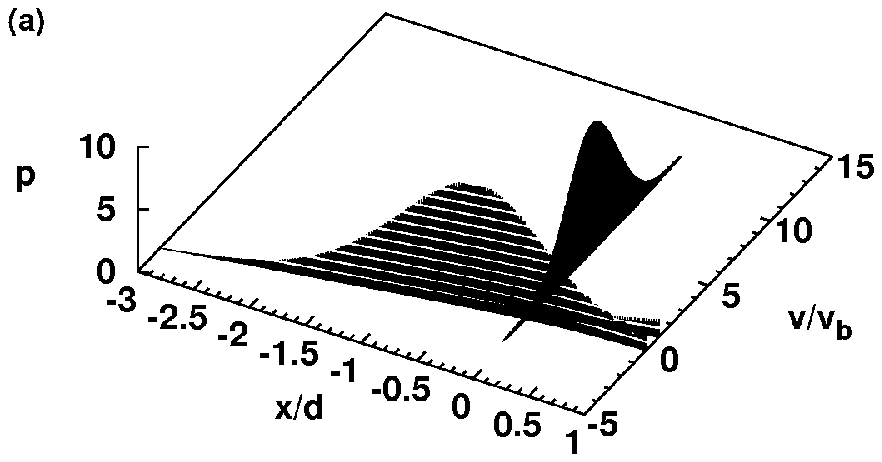}

\includegraphics[angle=0,width=0.90\linewidth]{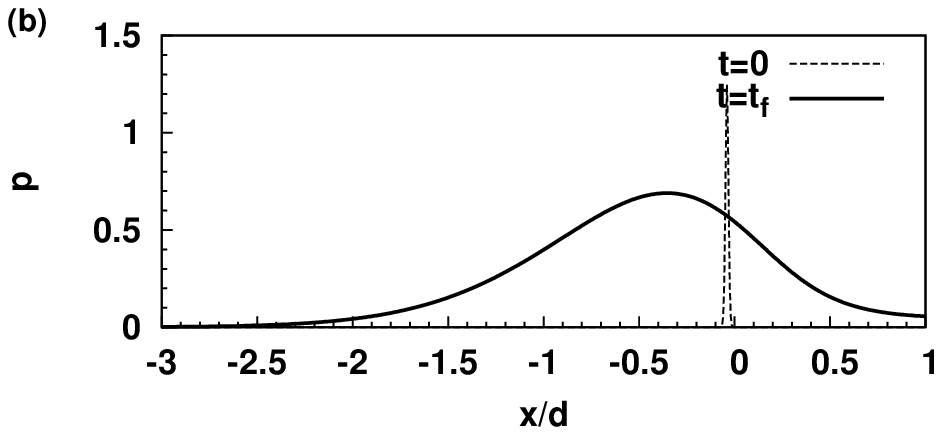}

\includegraphics[angle=0,width=0.90\linewidth]{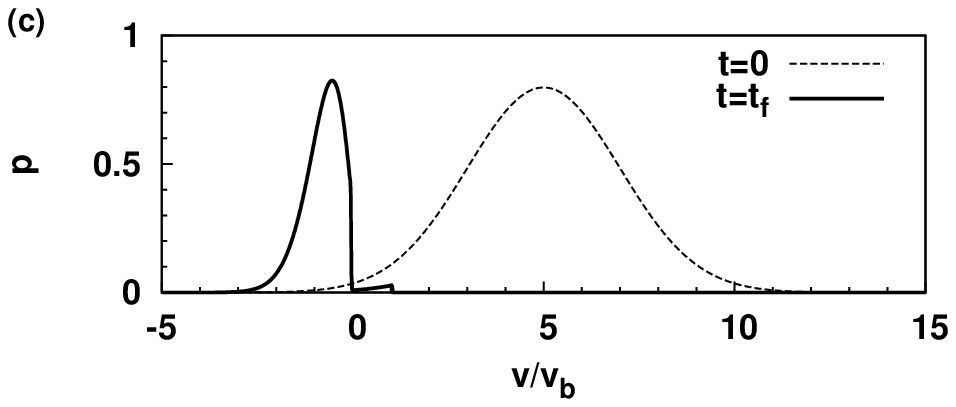}

\end{center}
\caption{\label{fig4} Classical setting,
$x_0/d = -0.04$, $\Delta x/d = 0.008$, $v_0/v_b = 5.0$ and $\Delta v/v_b = 2.0$;
(a) initial probability density $p_s$ (solid graph) and final
probability density $p_f$;
(b) $p_{x,s}$ ($t=0$, scaled
by a factor of $1/40$) and $p_{x,f}$ ($t=t_f$); (c) $p_{v,s}$ ($t=0$, scaled
by a factor of $4$) and $p_{x,f}$ ($t=t_f$).}
\end{figure}

\begin{figure}[t]
\begin{center}
\includegraphics[angle=0,width=0.90\linewidth]{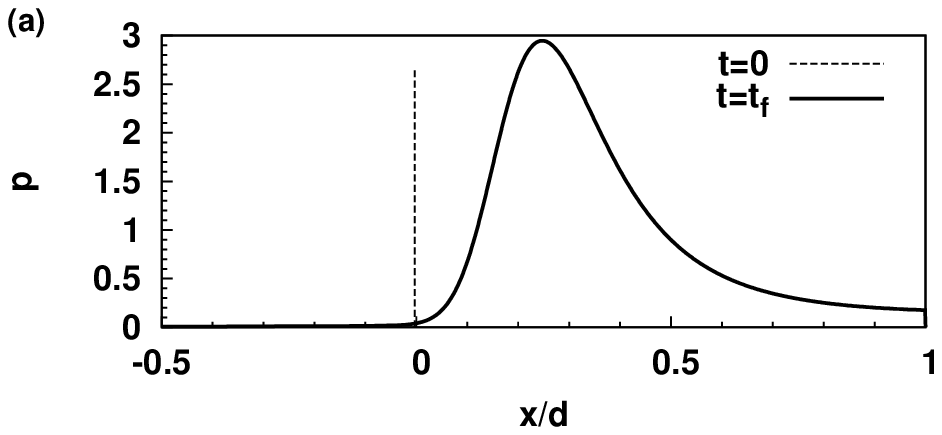}

\includegraphics[angle=0,width=0.90\linewidth]{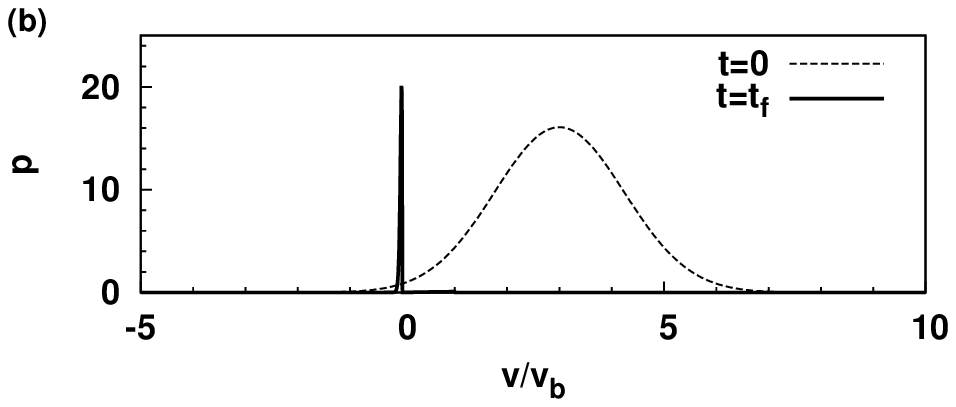}

\end{center}
\caption{\label{fig5} Classical setting,
$x_0/d = -0.003$, $\Delta x/d = 0.0003$, $v_0/v_b = 3.0$ and $\Delta v/v_b = 1.24$;
(a) $p_{x,s}$ ($t=0$, scaled
by a factor of $1/500$) and $p_{x,f}$ ($t=t_f$); (b) $p_{v,s}$ ($t=0$, scaled
by a factor of $50$) and $p_{x,f}$ ($t=t_f$).}
\end{figure}

The initial probability density $p_s$ for the parameters
$x_0/d = -0.04$, $\Delta x/d = 0.008$, $v_0/v_b = 5.0$ and $\Delta v/v_b = 2.0$
is shown in Fig. \ref{fig4}a (solid line). Also the final
probability density is shown in Fig. \ref{fig4}a (striped line).
The integrated probability densities
\begin{eqnarray*}
p_{x,s/f} (x) &=& \int dv \, p_{s/f} (x,v)\\
p_{v,s/f} (v) &=& \int dx \, p_{s/f} (x,v)\\
\end{eqnarray*}
for this example are shown in Fig. \ref{fig4}b and c.
Fig. \ref{fig4}c illustrates the slowing down and narrowing of the final velocity distribution (dashed line)
compared to the initial one (solid line). Moreover, 
a small fraction of particles have not hit the wall: 
they correspond to the final distribution in the
interval $0 < v/v_b < 1$.
One more example is shown in Fig. \ref{fig5}, where a rather drastic  
stopping can be seen. Note that the
decrease of the width of the velocity
distribution results always in an increase of the width of the position
distribution because the phase-space 
volume is conserved.
 
%
%
%
\section{Quantum setting\label{sect3}}
The dimensional Schr\"odinger equation for the particle and the moving mirror
potential is 
\begin{eqnarray*}
i \hbar \frac{\partial}{\partial t} \psi (x, t)
= - \frac{\hbar^2}{2 m} \frac{\partial^2}{\partial x^2}
\psi (x, t)
+ V \left(x - d \sqrt{t/t_f}\right) \psi (x, t),
\end{eqnarray*}
where $V(x)$ is the potential of the wall and the mass in the examples below 
is that of Rubidium ($m_{dim} = 14.19226\cdot 10^{-26}\, {\rm kg}$). 
The probability densities do not
depend on the mass of the particle classically, but in the quantum case they do.
The initial  wavefunction is a Gaussian, but not necessarily 
a minimum-uncertainty-product one,
\begin{widetext}
\begin{eqnarray*}
\psi_0 (x) = \frac{1}{N} \exp\left\{-\frac{\mu}{2 (1 + 2i \Delta v^2 \mu \delta)}
\left[i v_0 (\delta v_0 - 2 x) + 2 \Delta v^2 \mu ((x-\beta)^2 + 2 \delta v_0 \beta)\right]\right\},
\end{eqnarray*}
\end{widetext}
where
$\delta = \sqrt{4 \Delta x^2 - 1/(\Delta v^2 \mu^2)}/(2 \Delta v)$,
and $\beta = x_0 - \delta v_0$, $\mu = m/\hbar$.
The form of Heisenberg's uncertainty relation is 
$\Delta x \Delta v \ge 1/(2 \mu)$. 

In the calculations we shall use an infinitely high wall, i.e,
\begin{eqnarray*}
V_i (x) = \left\{\begin{array}{ll}\infty & : x \ge 0\\
0 & : x < 0
\end{array}\right.,
\end{eqnarray*}
as well as a more realistic Gaussian wall, with 
potential 
\begin{eqnarray*}
V_G (x) = V_0 \exp[-x^2/(2 \Delta x_V^2)].
\end{eqnarray*}
For cold atoms such a potential could be implemented by a detuned laser.

\begin{figure}[t]
\begin{center}
\includegraphics[angle=0,width=0.90\linewidth]{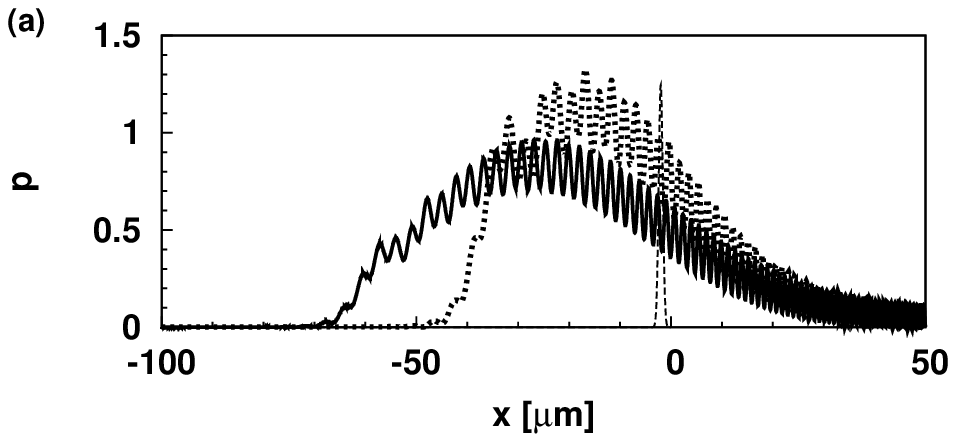}
\includegraphics[angle=0,width=0.90\linewidth]{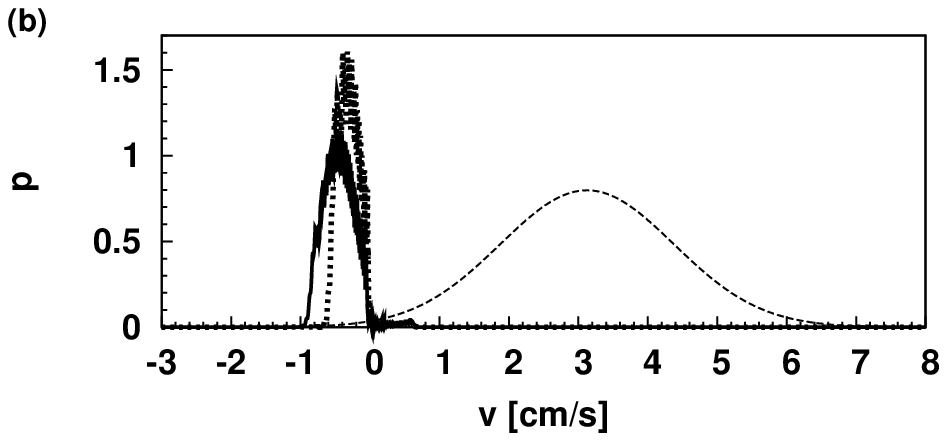}
\end{center}
\caption{\label{fig6} Quantum setting, see text for parameters;
(a) $p_{x,s}$ ($t=0$, dashed line, scaled
by a factor of $1/40$), $p_{x,f}$ with an ideal wall ($t=t_f$, thick solid
line), $p_{x,f}$ with a Gaussian wall ($t=t_f$, thick dotted line);
(b) $p_{v,s}$ ($t=0$: scaled
by a factor of $4$, dashed line),
$p_{v,f}$ with an ideal wall ($t=t_f$, thick solid
line), $p_{v,f}$ with a Gaussian wall ($t=t_f$, thick dotted line) and
$p_{x,f}$ ($t=1$).}
\end{figure}

The first example  corresponds to the parameters
\begin{eqnarray*}
x_0 &=& -2\, \mu{\rm m},\\
\Delta x &=& 0.4\, \mu{\rm m},\\
v_0 &=& 3.125\, {\rm cm/s},\\
\Delta v &=& 1.25\, {\rm cm/s}.
\end{eqnarray*}
We choose a boundary velocity $v_b = 0.625 {\rm cm/s} \ll v_0$
and $d = 50 \mu{\rm m}$, so that $t_f = 8\, {\rm ms}$.
We have $x_0/d = 0.04 \ll 1$. The height of the Gaussian potential
is set to $V_0/\hbar = 3.75\cdot 10^6/{\rm s}$ and its width is $\Delta x_V =
0.4 \mu{\rm m}$, both parameters must be chosen such
that the particle is reflected with high probability. 
The initial and final quantum probability densities are shown
in Fig. \ref{fig6}. Qualitatively, we see the same result in the quantum case as in the
classical case, with quantum interference fringes superimposed.

%

\begin{figure}[t]
\begin{center}
\includegraphics[angle=0,width=0.90\linewidth]{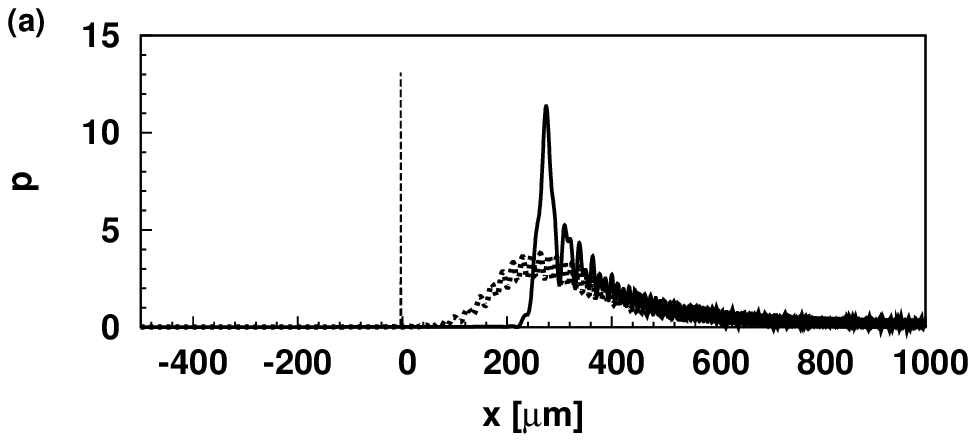}
\includegraphics[angle=0,width=0.90\linewidth]{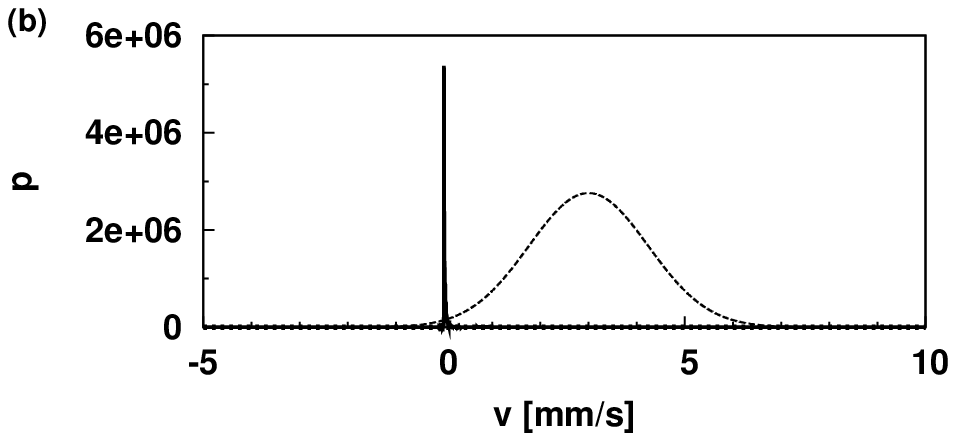}
\end{center}
\caption{\label{fig7} Quantum setting, see text for parameters;
(a) $p_{x,s}$ ($t=0$, dashed line, scaled
by a factor of $1/100$), $p_{x,f}$ with an ideal wall ($t=t_f$, thick solid
line), $p_{x,f}$ with a Gaussian wall ($t=t_f$, thick dotted line);
(b) $p_{v,s}$ ($t=0$: scaled
by a factor of $100$, dashed line),
$p_{v,f}$ with an ideal wall ($t=t_f$, thick solid
line), $p_{v,f}$ with a Gaussian wall ($t=t_f$, thick dotted line) and
$p_{x,f}$ ($t=1$).}
\end{figure}

Another spectacular example can be seen in Fig. \ref{fig7}, which shows the 
initial and final quantum probability densities for
\begin{eqnarray*}
x_0 &=& -3\, \mu{\rm m},\\
\Delta x &=& 0.3\, \mu{\rm m},\\
v_0 &=& 3\, {\rm mm/s},\\
\Delta v &=& 1.24\, {\rm mm/s},\\
d &=& 1000 \mu{\rm m},\\
t_f &=& 1\,{\rm s},\\
v_b &=& 1\, {\rm mm/s}.
\end{eqnarray*}
We have $v_b < v_0$ and $x_0/d = 0.003 \ll 1$.
The height of the Gaussian potential is set to
$V_0/\hbar = 3\cdot 10^4/{\rm s}$ and its width to $\Delta x_V = 0.8\, \mu{\rm m}$.

%
%
%
\section{Summary}
We have proposed a method to stop particles of unknown velocities or ensembles 
of particles with initial velocity spread by
reflecting them from an accelerated wall with trajectory $\sim \sqrt{t}$. 
For classical particles, the stopping is perfect if they are emitted 
from the origin at $t=0$. If the particle
starting point is not at the origin or if it is too slow, 
it is not perfectly stopped in a finite time.  
Explicit expressions are given for the final velocity; they also show
how to mitigate, and even suppress in a limit, the effect of non-ideal initial 
conditions.   
We may expect a similar behavior for a quantum wave packet and indeed,  
using numerical simulations with realistic and experimentally accessible
parameters, we have illustrated the efficiency of the method and discussed its
bounds.  
\section*{Acknowledgments}
We acknowledge A. del Campo for discussions and preliminary 
work. 
We acknowledge
``Acciones Integradas'' of the German
Academic Exchange Service (DAAD) \& Ministerio de
Educaci\'on y Ciencia;    
as well as Grants by Ministerio de Educaci\'on y Ciencia (FIS2006-10268-C03-01)
and Basque Country University UPV-EHU (GIU07/40).  
SS and AR acknowledge support by the German Research Foundation (DFG).

\end{document}